# SAMPAI: A VR Framework for Industrial Safety Training Inspired by Cultural Heritage Education

**Gianni Vercelli**[1]
DIBRIS -University of Genoa, Via alla Opera Pia, 13, Genoa, 16145, GE, Italy

**Saverio Iacono**[2]
DIBRIS - University of Genoa, Via alla Opera Pia, 13, Genoa, 16145, GE, Italy

**Luca Martini**[3]
DLCM – University of Genova, P.za S.Sabina, 2, 16124, Genova Italy;

**Michele Zardetto**[4]
DIBRIS -University of Genoa, Via alla Opera Pia, 13, Genoa, 16145, GE, Italy

**Daniele Zolezzi**[5]
DIBRIS -University of Genoa, Via alla Opera Pia, 13, Genoa, 16145, GE, Italy

January 29, 2026

## Abstract

This study explores the application of Virtual Reality (VR) to industrial safety training by adapting immersive design principles from cultural heritage education. The SAMPAI simulator, developed for the IPLOM refinery in Busalla (Italy), offers a controlled environment that enhances spatial awareness and procedural retention in high-risk scenarios. The project emphasizes user-centered interaction and anticipates future integration with Augmented Reality (AR) for real-time operational support. To ensure reliability, a rigorous validation protocol is proposed: initial testing with a small operator group will identify usability issues, followed by incremental expansion to strengthen system robustness and real-world relevance. By shifting training from abstract instruction to situated, experiential learning, SAMPAI bridges humanistic and technical domains. The result is a scalable, immersive framework that enhances readiness and supports safe, informed decision-making in complex industrial environments.

## 1. Introduction

In the context of Industry 4.0 and Cyber-Physical Systems (CPS), Extended Reality (XR) technologies—particularly Virtual Reality (VR)—are redefining training paradigms by offering immersive experiences that boost engagement and retention of complex content. In the cultural sector, numerous studies have shown that VR applied to the teaching of historical and artistic heritage enables the creation of immersive narratives, fosters empathy toward the studied contexts, and enhances experiential learning through high-fidelity 3D reconstructions [1]. Simultaneously, VR has demonstrated superior effectiveness compared to traditional methods in industrial safety training—especially in hazard recognition, emergency management, and evacuation procedures: meta-analyses and systematic reviews report significant improvements in response times and retention of operational protocols, thanks to the engagement afforded by virtual environments [2]. Practices developed in heritage education—such as the integration of gamification mechanics, VR serious game, the use of immersive storytelling, and a user-centered approach based on immediate feedback—can serve as a valuable methodological model for training in high-risk industrial contexts [3]. In particular, trials of "VR games" in cultural sites have shown that gamified exploration and progressive objectives strengthen intrinsic motivation and active learning capabilities [4].

[1] gianni.vercelli@unige.it
[2] saverio.iacono@unige.it
[3] luca.martini@edu.unige.it
[4] s5719958@studenti.unige.it and mich.zardetto@gmail.com
[5] daniele.zolezzi@edu.unige.it

In this contribution, we propose a modular VR Edu-Training framework that, building on the experience gained in cultural heritage—where the humanities guide narrative and interactive design—extends to industrial safety training. The prototype, developed using Unreal Engine and Meta Quest devices, integrates hazardous-emission drills and evacuation simulations, ensuring scalability toward future AR extensions and interoperability with Metaverse architectures for large-scale training [5]. The following chapters will examine the state of the art of VR applications in safety training, describe the design methodology of the SAMPAI prototype, and present proposals for validation sessions.

## 2. Related Works

Previous studies have demonstrated the potential of virtual and augmented reality in improving operator safety during training [6] taking in account the necessity to avoid excessive stimuli could interfere with effective learning [7]. In the steel industry, the utility of VR has been demonstrated for visualizing plant-wide processes, improving both operational optimization and workforce training. This approach shows how VR can make complex metallurgical workflows more accessible and intuitive for trainees [8]. Similarly, advancements in training methodologies were achieved by integrating AR into Operator Training Simulators (OTS). This approach fostered better collaboration between field and control room operators during chemical accidents, leading to a significant improvement in training outcomes [9]. Early applications, such as the omVR system, showcased the potential of VR to simulate high-risk scenarios, including hydrogen sulfide emissions in petroleum refineries. This system provided a safe and immersive environment for training personnel in emergency responses, demonstrating the advantages of Head-Mounted Displays (HMDs) in enhancing engagement and realism. However, limitations in real-time integration and scalability across diverse industrial contexts hindered its broader adoption at the time [10]. Building on these advancements, a VR-enabled system was introduced for real-time equipment monitoring, integrating industrial sensors with immersive environments through MQTT protocols. This approach enhances situational awareness and reduces the cognitive load on operators by providing real-time visualizations of equipment health [11].

VR has also been successfully applied in industrial and academic training. For instance, a research published in Education for Chemical Engineers examined VR's role in education and training for chemical and biochemical engineering. The study concluded that VR provides an innovative approach to simulate complex experiments and processes, enhancing understanding and ensuring safety for both students and professionals [12]. A particularly noteworthy case for our research is the development of a VR-based safety training prototype designed for oil refinery operators. This system enables personnel to engage with realistic emergency scenarios within a secure virtual environment, as highlighted in recent studies. Such an approach not only minimizes risks to operators but also significantly enhances their readiness to handle real-life emergencies. These case studies were selected for their direct relevance to our investigation into improving safety and operational efficiency in oil refineries through VR applications. They illustrate VR's potential not only for real-time monitoring and predictive maintenance but also for training and emergency management. Such innovative solutions offer significant benefits in enhancing safety and reducing risks in complex industrial environments, providing a strong foundation for further exploration in our research [10]. In a different context, yet highly relevant to "stress management," the use of VR was explored for communication training, particularly for police officers navigating high-stress scenarios. By prioritizing realism, repetition, and feedback, their work showcased the value of VR in skill development under pressure. Though effective for law enforcement, the potential for such methods to improve emergency communication in industrial settings has yet to be fully explored [13].

The transformative role of extended reality (XR) technologies in education has been emphasized, with a focus on their potential to better prepare students for complex real-world scenarios. These technologies enable immersive, participatory training environments that enhance critical decision-making and facilitate skill acquisition [14]. Meanwhile, a study explored the usability of VR and AR interfaces for developing interactive simulations, demonstrating that these technologies can surpass traditional desktop methods in terms of immersion and user engagement [15]. While XR technologies offer immense benefits for professional training, education, and safety, their adoption is not without challenges. Ethical concerns, including data privacy, potential addiction, and psychological impacts, remain critical barriers, particularly in high-stakes industries. These issues underscore the need for responsible implementation strategies that balance innovation with safety and ethical considerations. These issues are especially pertinent in high-stakes industries, where extended use of immersive environments may lead to unintended consequences [16]. Overall, the literature underscores VR and AR's capacity to revolutionize training and operational readiness by providing realistic, interactive, and risk-free environments.

## 3. Material and Methods

This research aims to explore novel applications of AR and VR technologies in industrial settings, with a particular focus on their potential to enhance collaboration, operational efficiency, and safety in complex environments such as oil refineries. The study identifies four specific objectives to systematically evaluate these immersive technologies and their impact on critical industrial processes:

- Evaluating AR for Real-Time Collaboration: The objective is to explore how augmented reality can enhance communication and teamwork among refinery operators. By enabling shared, real-time visualization of critical data, AR is expected to support decision-making and improve coordination. The study assesses communication quality and efficiency when AR is embedded into operational workflows.
- Analyzing VR in Training and Emergency Management: The focus lies on the use of virtual reality for training, especially in emergency response scenarios. Immersive simulations aim to increase preparedness while reducing real-world risks. Operator performance will be compared between VR-based and traditional training methods, with expectations of better retention and crisis management through VR.
- Integrating IoT with VR/AR for Remote Monitoring: Combining IoT data with immersive technologies such as VR and AR could offer a more complete view of refinery operations. Real-time access to sensor data is anticipated to improve monitoring capabilities and response times. The research evaluates usability and effectiveness in enhancing situational awareness and control.
- Assessing Immersive Technologies in Ongoing Training: AR and VR are investigated as tools for continuous operator training, particularly for updating safety procedures and introducing new technologies. Immersive approaches are hypothesized to increase engagement and retention, helping operators stay prepared for evolving industrial challenges. Key metrics include long-term retention and user feedback.

By addressing these objectives, this research seeks to establish a framework for the effective adoption of AR and VR technologies in industrial environments, demonstrating their potential to transform key operational and safety processes. The findings are expected to contribute to the growing body of knowledge on immersive technologies and their practical applications in high-stakes industrial contexts. The following research hypotheses address a specific dimension of the potential impact of AR and VR technologies in industrial environments:

1. H1: the use of Augmented Reality for real-time collaboration among operators enhances communication and coordination, reducing operational errors and improving decision-making efficiency.
2. H2: Virtual Reality provides a significant advantage in operator training for emergency scenarios. Operators trained using VR are hypothesized to exhibit improved emergency management skills, including faster response times and greater accuracy in implementing safety procedures.
3. H3: integrating IoT data with VR/AR applications enhances remote monitoring of industrial plants and improves incident response capabilities by enabling real-time visualization of operational data.

4. H4: the use of immersive technologies such as AR and VR in continuous operator training increases learning effectiveness and information retention compared to traditional training methods.

These hypotheses will guide the design and implementation of the research methodology, enabling a systematic evaluation of the benefits and challenges associated with integrating AR and VR technologies into industrial workflows.

The methodology in this project involved the comprehensive development and deployment of a VR training environment aimed at enhancing safety protocols and operational readiness specifically for industrial waste monitoring at the IPLOM refinery. Given the high-stakes nature of refinery operations, the simulation needed to achieve high realism and fidelity, capturing both the spatial and procedural complexities inherent to such environments. The VR simulation is designed to provide operators with an immersive, controlled experience in which they can practice responding to potential hazards without being exposed to actual risks. To meet these requirements, Unreal Engine served as the main platform for the simulation, chosen for its advanced 3D rendering capabilities, flexible physics engine, and extensive VR support. These capabilities allowed the project team to create an environment that replicates not only the appearance but also the interaction and responsiveness of real-world industrial settings, where precision and timely reactions are critical. Additionally, OpenXR standards were integrated to maintain cross-platform compatibility, which ensures the simulation's adaptability to various VR and AR devices, making it a robust solution that can evolve with technological advancements in the field. The hardware selected for the VR environment included Meta Quest 2 and Meta Quest 3 headsets, which were specifically chosen for their portability, wireless operation, and ease of use. These devices are all-in-one systems that do not require an external computer connection, making them suitable for scalable implementation across multiple sites without needing extensive setup. The Meta Quest 3, with its advanced hand-tracking and built-in cameras, is especially relevant as it supports AR functionality, enabling potential future applications of AR in training and maintenance. The headset's capabilities allow seamless transitions between immersive VR experiences and AR overlays, providing a versatile training tool that can support operators in both virtual and real-world settings. For instance, with AR capabilities, real-time information such as equipment performance data or safety alerts could be overlaid on machinery during routine operations, further enhancing situational awareness and decision-making in on-site tasks. Meta Quest headsets feature integrated capabilities for session recording and real-time streaming to any PC equipped with a browser, providing additional flexibility for monitoring and analysis during training or operational activities. In constructing the VR simulation, the team utilized elements from Unreal Engine's "Factory Environment Collection" to create a virtual office space modeled closely after a real industrial workspace. This space serves as the central control hub within the simulation, equipped with various interactive elements that allow users to familiarize themselves with safety protocols and critical response tasks. The office environment includes a highly detailed 3D map of the refinery, providing a comprehensive overview of the facility's layout. This interactive map enables users to visualize and practice navigating the refinery's spatial structure, including emergency exits and key operational zones. By integrating these elements into the VR experience, the simulation provides a realistic and structured environment where users can engage in training scenarios, practicing decision-making skills under simulated stress. Functional items within the office, such as virtual tablets, replicate alarm systems and monitoring stations, allowing operators to interact with control systems as they would in actual refinery settings, thereby fostering a sense of procedural familiarity.

An essential component of the VR simulation was to provide users with realistic, intuitive interaction capabilities that replicate handling and operational tasks found in the refinery. Using Unreal Blueprint visual scripting system, the team developed custom interaction mechanics suited to training needs, allowing for a dynamic, immersive experience. For example, the Blueprint GrabType feature was utilized to create lifelike object manipulation, enabling users to "pick up" and handle virtual equipment, like real-world refinery tools. Additionally, a custom interaction called Blueprint Remote Pulling was developed to enable users to attract objects from a distance toward the VR operator. This functionality is particularly useful for scenarios where operators need to retrieve tools or items without physically approaching them, although it does not yet simulate direct interactions with machinery or controls. These interaction styles ensure that the simulation covers a broad spectrum of training needs, from hands-on equipment management to remote operation scenarios, making it an effective tool for various skill-building exercises within industrial settings. To fully address safety training requirements, the VR simulation includes a range of scenarios that model emergency situations operators might encounter, such as gas leaks or equipment malfunctions. In these scenarios, VR-generated alerts guide users through safety protocols, including the use of multi-gas detectors and proper personal protective equipment (PPE). Additionally, the simulation provides users with access to detailed 3D virtual maps of the IPLOM refinery, enabling them to view evacuation routes clearly. However, at this stage of development, it does not allow for virtualwalkthroughs or active practice of these routes within the simulation itself. This hands-on approach helps solidify critical response skills in a safe and controlled environment, enabling operators to experience and manage simulated emergencies. By providing this immersive training, operators can improve their response time and familiarity with safety protocols under semi-realistic stress conditions, an invaluable asset in the hazardous industrial context of the refinery.
To ensure accessibility for all users, especially those with no prior experience with VR technologies, the project includes a dedicated preliminary VR training course. This introductory module, developed as a separate level in Unreal Engine, is designed to acquaint beginner refinery operators with the VR interface and interaction mechanics. It emphasizes:

- Basic Navigation: guiding users on how to move within the virtual environment using controllers or hand gestures, based on the capabilities of the Meta Quest 2 and 3 headsets.
- Interacting with Objects: demonstrating how to manipulate objects, such as virtual tools or controls, using features like the Blueprint GrabType and Remote Pulling functionalities.
- Scenario Walkthroughs: a step-by-step walkthrough of simplified emergency scenarios to build initial confidence.

By integrating this beginner-level course, the simulation ensures that all operators, regardless of their technical expertise, can transition smoothly into more advanced training sessions. This approach minimizes the learning curve associated with immersive technologies and promotes effective engagement across diverse user groups.
Future adaptability was a central consideration in the design of this VR simulation. With Meta Quest 3's AR capabilities, the system is built to support future integration with AR for enhanced training and on-site maintenance assistance. For example, the AR functionality could allow operators to view real-time data overlays directly on refinery equipment, supporting instant access to operational metrics and safety information. Such integration would enable a blended learning experience, where operators can train both in virtual and real-world contexts, transitioning seamlessly between VR and AR. This level of adaptability positions the project as a forward-thinking solution that can evolve alongside advancements in XR technology, meeting the long-term training and operational needs of high-risk industrial environments. .

## 4. SAMPAI Project

The SAMPAI simulator was developed for the IPLOM refinery in Busalla to offer an immersive and secure environment where personnel can acquire and enhance their skills in managing risks associated with heavy industry operations. The name SAMPAI, an acronym for "Simulazione di Addestramento e Monitoraggio ai Pericoli dell'Ambiente Industriale" in Italian ("Simulation for Training and Monitoring of In-dustrial Environment Hazards" in English), underscores the simulator's emphasis on hazard prevention and management, while also facilitating the monitoring of critical refinery sensors to ensure optimal safety.
The VR training simulation developed for the IPLOM refinery aims to create a controlled, immersive training environment that mirrors the complexity and precision of the safety protocols necessary in hazardous industrial settings. The hands-on, immersive nature of the VR experience could enhance procedural retention, as users could engage in repeated practice of emergency response scenarios. By offering a realistic yet controlled setting, the VR training removes the physical risks and logistical constraints associated with real-life drills, making it possible for operators to rehearse responses without interruption or resource limitations. This freedom to practice safely and repeatedly contributes to better recall of critical procedures, as operators internalize each action through active simulation [17].
This virtual environment provides a highly realistic experience where operators can engage in rigorous training on a variety of emergency response procedures. One of the most impactful achievements of this VR application is its capacity to present high-risk scenarios, such as hydrogen sulfide ($H_2S$) gas leaks, with accuracy and immediacy, but without exposing trainees to the physical dangers of actual leaks. The system simulates real-world alarm triggers and responses, incorporating audio and visual cues that guide users through each step of the emergency protocols. Operators practice critical decision-making under simulated pressure, reinforcing their ability to respond quickly and effectively in real emergencies. This VR training replicates the use of essential safety tools, including multi-gas detectors, which alert users to dangerous gas levels, and emergency escape masks, which they learn to do correctly in response to simulated toxic releases (Figure 1).

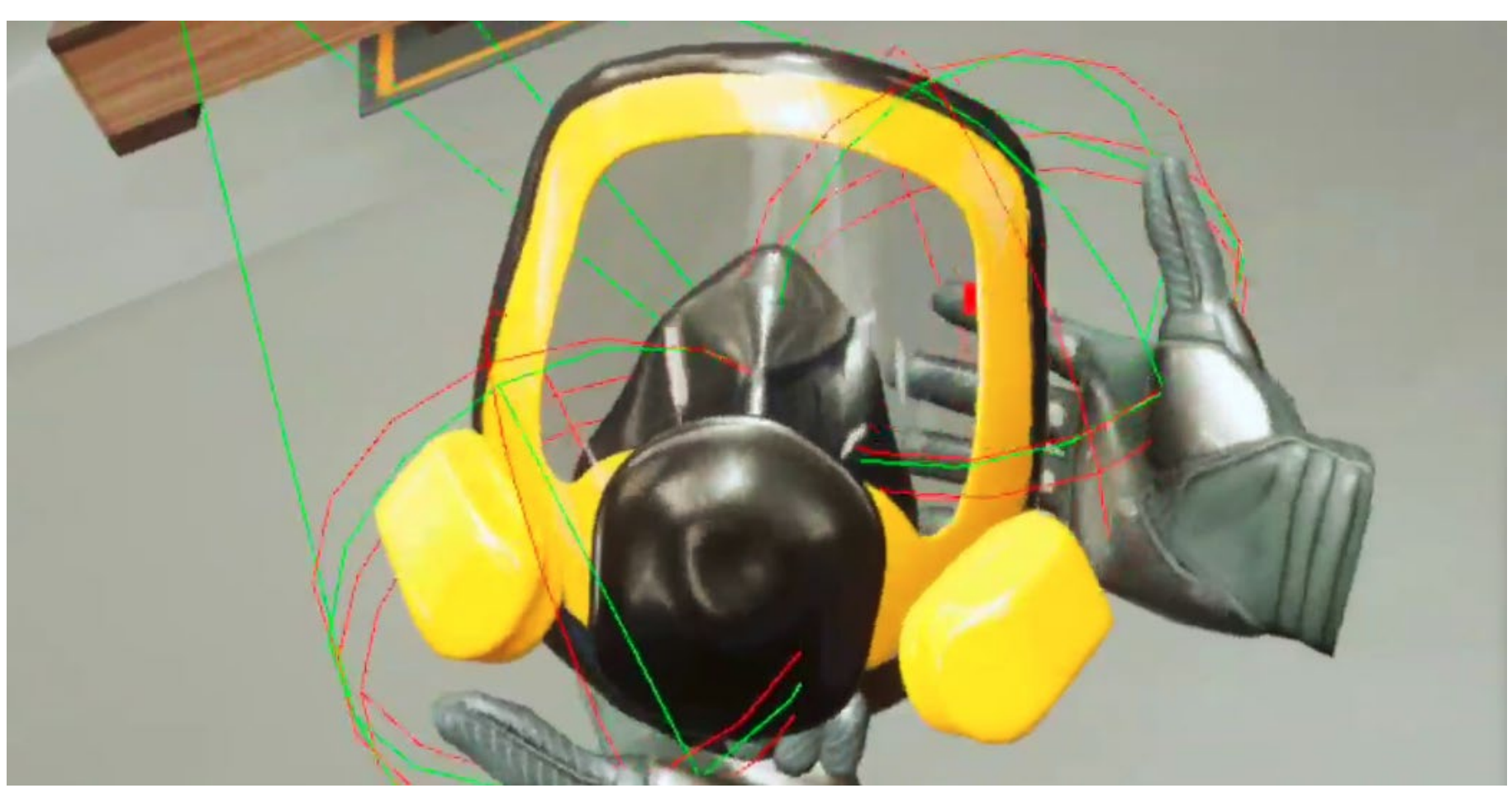

Figure 1 - Example of an essential tool to help simulate safety.

These devices are virtually manipulated to match real handling and operation, enabling users to familiarize themselves with equipment functionality in an environment that mimics physical conditions. Training is segmented into three structured courses, each lasting approximately 5-10 minutes, allowing users to engage repeatedly with key safety protocols. The courses cover core routines—such as daily checks, hazard recognition, emergency equipment use, and structured evacuation procedures. Through direct interaction within the VR setting, operators gain hands-on experience that strengthens procedural memory, response speed, and overall confidence in navigating high-stress, high-risk situations specific to the refinery's operational context. A critical element of the VR environment is the integration of a detailed 3D map that replicates the refinery layout, serving as both a spatial reference and an interactive training tool for emergency preparedness (Figure 2).

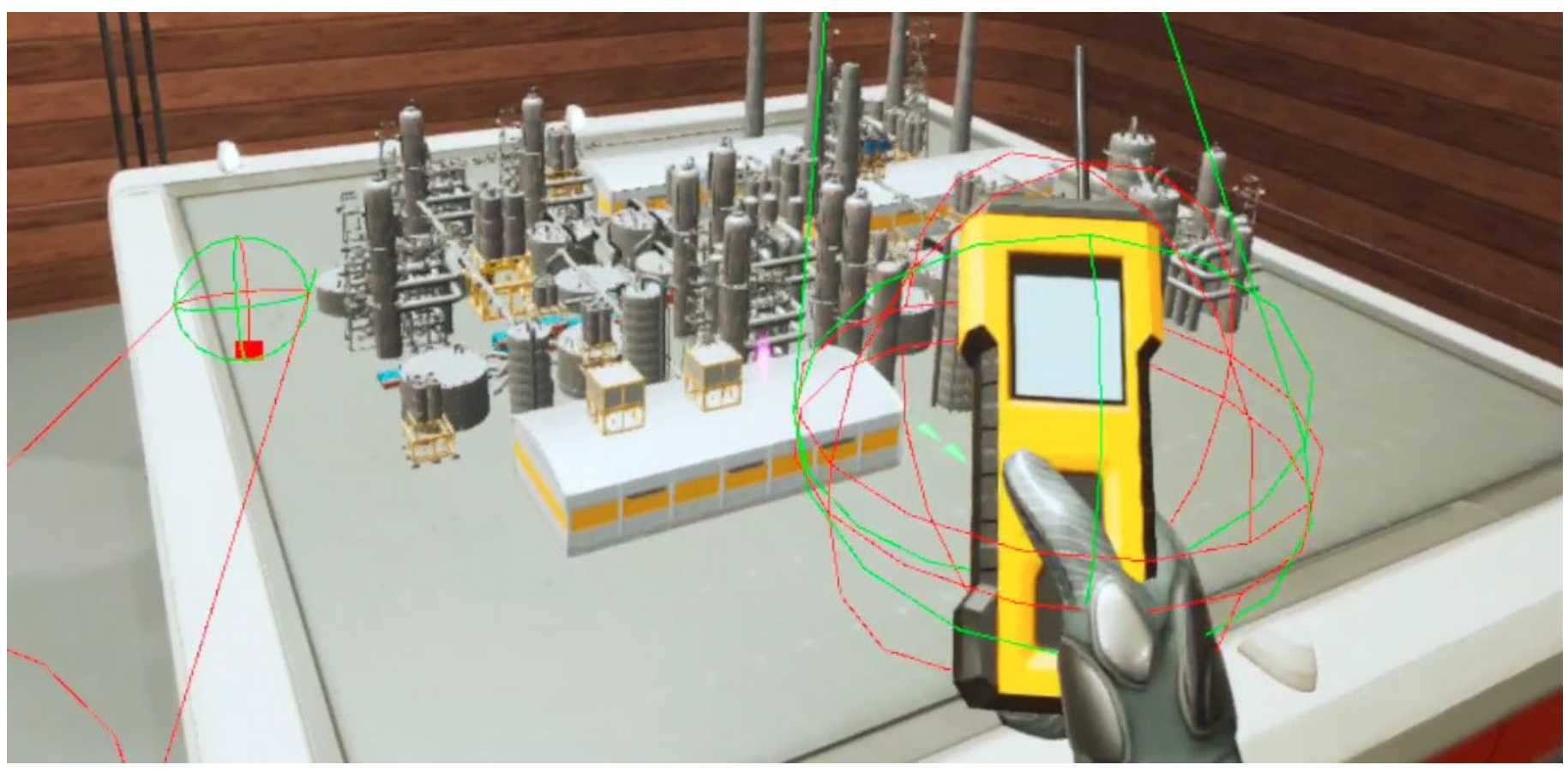

Figure 2 - Example of an essential tool to help simulate safety.

Although simplified for this prototype, the map effectively captures the key zones of the refinery, enabling operators to practice safe navigation through various hazardous areas, including storage tanks, pipelines, and designated safety zones. The virtual map is fully integrated into training scenarios, allowing users to simulate real-time responses to emergency signals. For example, the map displays animated indicators, such as red flame icons or gas cloud symbols, to mark specific locations affected by simulated incidents like leaks or fires. These visual cues are carefully designed to prompt immediate situational awareness, guiding trainees to recognize danger zones and prioritize safe evacuation routes.

Through these interactive elements, the VR system immerses users in situational awareness exercises, where they must evaluate and choose safe paths based on the unfolding scenario, reinforcing both spatial memory and response precision. In larger and complex industrial settings like IPLOM's refinery, where emergency exits, safety stations, and evacuation routes are spread across extensive areas, this realistic, hands-on navigation training is crucial. By using the map to rehearse the emergency pathway, operators can internalize the refinery's layout and develop a stronger familiarity with critical routes, minimizing potential confusion and reaction times in real-life incidents. This feature provides an invaluable tool for training operators to navigate under pressure, instilling a level of readiness that aligns closely with the demands of actual high-risk industrial plants. In a future implementation of AR, it will be possible to visualize in real-time how evacuation routes change dynamically based on contingent emergencies or blocked paths.

The simulation's interactive features, developed through Unreal Engine's Blueprint system, are pivotal in preparing users for real-world equipment handling and safety protocols. By using the Blueprint GrabType function, the simulation allows trainees to virtually "pick up" and operate essential tools like multi-gas detectors, fire extinguishers, and protective masks, closely replicating the tactile experience of handling these devices. This function offers realistic object interactions, where users can grasp, manipulate, and position virtual tools, practicing the fine motor skills and precise actions necessary for safe operation. For instance, operators can practice checking and calibrating a gas detector, simulating the exact hand movements required to use the device under urgent conditions. Additionally, the Blueprint Remote Pulling function enables users to engage with equipment that may be positioned outside of their immediate reach. This feature is particularly valuable in industrial contexts, where operators frequently need to control machinery or activate safety switches from a distance, such as during hazardous material handling or in restricted zones. The combination of realistic equipment handling and remote access allows operators to practice essential, routine tasks in an environment that mirrors the physical constraints of the refinery, bridging the gap between virtual practice and actual equipment use. By providing this seamless transition to real-world applications, the VR simulation equips operators with the procedural familiarity and confidence needed to handle equipment effectively and safely in high-stakes, on-site conditions..

The VR simulation developed for IPLOM's training program showcased the transformative potential of immersive technology in industrial safety training. By providing a controlled, risk-free environment, the VR system allows operators to practice critical safety protocols and emergency responses with high accuracy and repeated exposure (Figure 4). This approach not only enhances trainees' skills but also enables significant reductions in training costs, as it minimizes the need for frequent physical drills and reduces the wear on equipment used for training. Additionally, the immersive VR experience contributes to faster response times and improved procedural retention, as operators can refine their reactions to emergency scenarios in a realistic but safe set-ting.

Beyond immediate benefits, this VR model is highly scalable and adaptable, providing a versatile training template that can be replicated in other industrial facilities with similar environmental and operational hazards. The model's adaptability also al-lows for customizations to meet specific site requirements, ensuring that each implementation can reflect the unique safety and operational protocols of different industrial contexts. This flexibility makes the VR training simulation not only a valuable tool for IPLOM but also a promising advancement in the field of industrial safety education, with the potential to become a standard solution for high-risk industries. By improving readiness and lowering operational risks, the VR system offers a substantial advantage in building a safer, more responsive workforce.

## 5. Future Works

The VR platform developed for IPLOM's training program demonstrates the transformative potential of immersive technologies in industrial safety (Figure 3). Its flexibility enables adaptive training schedules and site-specific customization, allowing personnel to revisit and refine essential skills—an approach shown to reduce training costs and improve operational readiness in refinery environments [2].

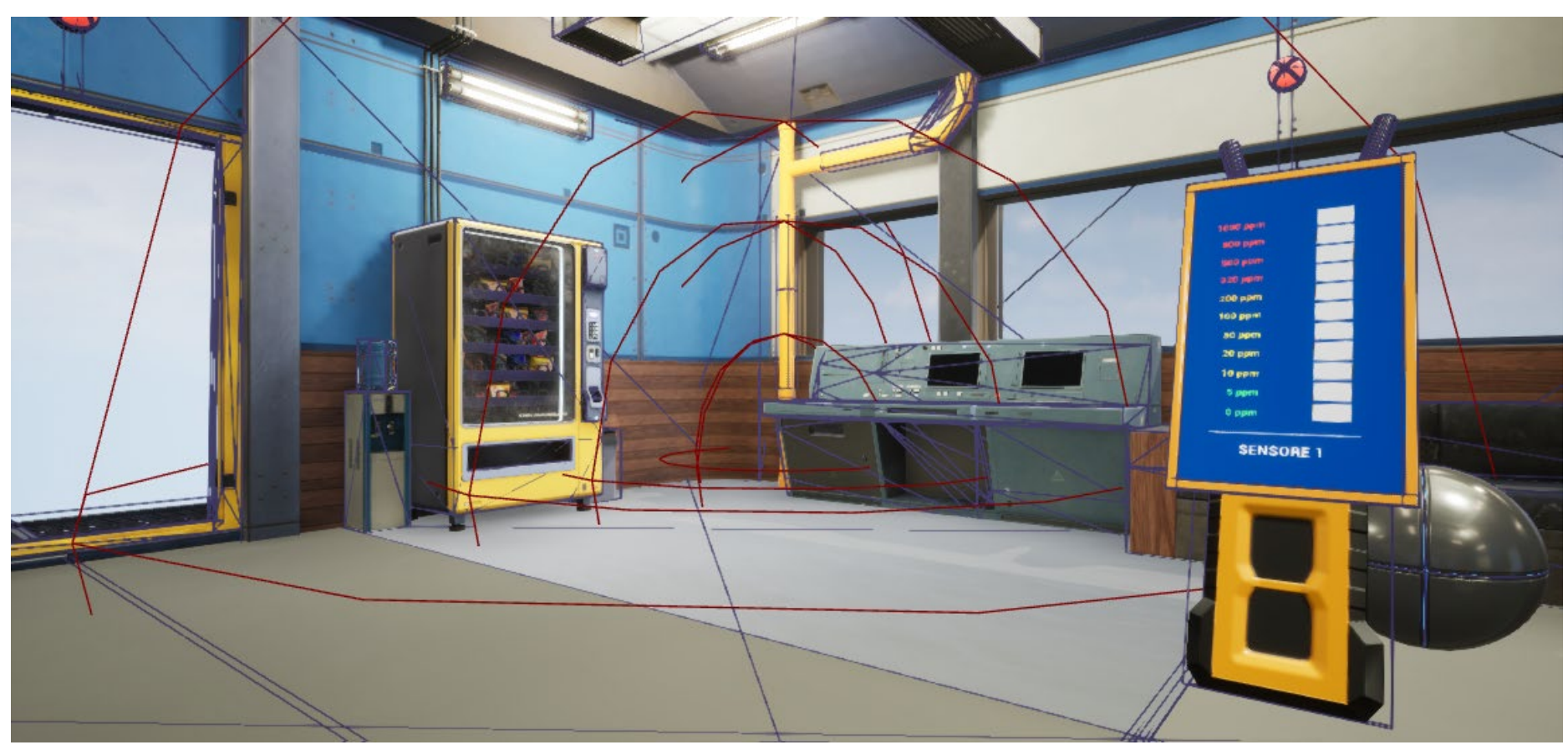

Figure 3 - SAMPAI's safe learning environment example.

The system's design also anticipates future integration with AR technologies. Leveraging the Meta Quest 3 headset's capabilities, the system could provide real-time data overlays—such as temperature, pressure, and operational status—directly onto equipment. This would turn the headset into a dynamic tool for both training and operational support. For example, operators could receive step-by-step maintenance instructions or immediate performance feedback superimposed on a machine, enhancing situational awareness and enabling precise, data-driven responses in the field.

Preliminary tests conducted in the early stages of the project yielded encouraging feedback regarding system usability and engagement. To substantiate these findings scientifically, we are evaluating the following multi-phase validation methodology. Initial testing will involve a controlled group of operators to identify usability issues and refine the experience. Subsequent phases will incrementally expand the participant pool, improving system robustness and ensuring that training scenarios reflect real-world operational demands. This methodical approach will also provide critical insights into user behavior, system performance, and procedural adherence—essential for planning effective AR integration.

An experimental protocol will compare task execution time and accuracy between VR and traditional training groups [2]. Spatial awareness will be assessed using the mental rotations test, a validated tool in high-risk cognitive domains [18]. Retention will be measured at 7 and 28 days post-training, consistent with meta-analytic findings that show immersive platforms offer significantly higher retention rates than conventional methods [19]. Ultimately, by bridging simulation and real-world application, the integrated VR/AR system promises to elevate safety standards, enhance decision-making, and support a more responsive and skilled industrial workforce.

The SAMPAI framework illustrates a broader convergence between industrial training and core concerns of Cyber Humanities. Virtual environments such as those developed in this project not only support procedural skill acquisition but also reconfigure how knowledge is embodied, contextualized, and accessed. Rather than acting solely as technical systems, VR and AR platforms mediate experience by embedding users within operational simulations where perception, spatial reasoning, and decision-making are actively engaged. The shift from abstract instruction to situated, sensorimotor learning reflects a transformation in the epistemology of training. From this perspective, immersive training becomes a site where technical expertise intersects with human-centered design, highlighting the cultural and experiential dimensions of industrial knowledge. To further enhance long-term retention and user engagement, the integration of narrative elements is currently under evaluation, with the goal of evolving the simulation into a serious game that fosters active participation and immersive learning.